\documentclass[a4paper,fleqn,usenatbib]{mnras}

\usepackage{mathptmx}

\usepackage[T1]{fontenc}
\usepackage{ae,aecompl}

\usepackage{graphicx}	
\usepackage{amsmath}	
\usepackage{amssymb}	

\title[Changing-look AGN explained by state transition]
{Explaining changing-look AGN with state transition triggered by rapid
mass accretion rate drop
}

\author[H.~Noda \& C.~Done]{
Hirofumi Noda$^{1,2,3}$\thanks{E-mail: hirofumi.noda@astr.tohoku.ac.jp}
and Chris Done$^{3}$
\\
$^{1}$ Frontier Research Institute for Interdisciplinary Sciences, Tohoku University,  6-3 Aramakiazaaoba, Aoba-ku, Sendai, Miyagi 980-8578, Japan\\
$^{2}$ Astronomical Institute, Tohoku University, 6-3 Aramakiazaaoba, Aoba-ku, Sendai, Miyagi 980-8578, Japan\\
$^{3}$ Centre for Extragalactic Astronomy, Department of Physics, University of Durham, South Road, Durham DH1 3LE, UK
}

\date{Accepted 2018 July 25. Received 2018 July 20; in original form 2018 May 20}

\pubyear{2015}

\begin{document}
\label{firstpage}
\pagerange{\pageref{firstpage}--\pageref{lastpage}}
\maketitle

\begin{abstract}
We model the broadband (optical/UV and X-ray) continuum spectrum of
the ``changing-look''  active galactic nucleus (AGN) Mrk~1018 as it fades 
from Seyfert 1 to 1.9 in $\sim 8$~years. The brightest spectrum, with
Eddington ratio $L/L_{\rm Edd}\sim 0.08$, has a typical type 1 AGN
continuum, with a strong ``soft X-ray excess'' spanning between the UV 
and soft X-rays. The dimmest spectrum, at $L/L_{\rm Edd}\sim 0.006$, 
is very different in shape as well as luminosity, with the soft excess
dropping by much more than the hard X-rays. The soft X-ray excess 
produces most of the ionizing photons, so its dramatic drop leads to the 
disappearance of the broad line region, driving the ``changing-look'' 
phenomena. This spectral hardening appears similar to the soft-to-hard 
state transition in black hole binaries at $L/L_{\rm Edd} \sim 0.02$, 
where the inner disc evaporates into an advection dominated
accretion flow, while the overall drop in luminosity appears consistent 
with the Hydrogen ionization disc instability. Nonetheless, both processes 
happen much faster in Mrk~1018 than predicted by disc theory. We critically 
examine scaling from galactic binary systems, and show that a major difference 
is that radiation pressure should be much more important in AGNs, so that 
the sound speed is much faster than expected from the gas temperature. 
Including magnetic pressure to stabilize the disc shortens the timescales 
even further. We suggest that all changing-look AGNs are similarly associated
with the state transition at $L/L_{\rm Edd} \sim$ a few percent.

\end{abstract}




\section{Introduction}

The optical variability of active galactic nuclei (AGNs) is puzzling.
The viscous timescale for changing the mass accretion rate onto a
supermassive black hole (SMBH) is extremely long for a standard thin
disc at radii corresponding to those which produce the optical emission, 
at least thousands of years for typical AGN parameters. Yet AGNs typically 
show stochastic optical variability of a factor of a few on timescales of 
months--years
(e.g., \citealt{2009ApJ...698..895K}; \citealt{2010ApJ...721.1014M}).
This flux change could instead be produced from reprocessing of the
X-ray emission from a central compact source. An X-ray tail
accompanies the disc emission in all AGNs (e.g., \citealt{1994ApJS...95....1E};
\citealt{2016ApJ...819..154L}) requiring that some of the gravitational power
is dissipated in optically thin material rather than in the optically
thick, geometrically thin disc. These X-rays vary on rapid timescales,
indicating that they come from a compact region around the SMBH, 
and the fast variable X-ray heating can result in fast optical variability
\citep{1992ApJ...393..113C}. However, while reprocessing can provide 
an overall explanation of the fast timescales, reverberation studies of 
the optical continuum imply size scales which are generally a factor of 
a few larger than the thin disc predictions (e.g.,
\citealt{2014ApJ...788...48S}; \citealt{2016MNRAS.456.4040T};
\citealt{2016ApJ...828...78N}), and detailed models which use the
observed X-ray emission to predict the optical flux do not match well
to the observed light curves (e.g., \citealt{2017MNRAS.470.3591G}).

Nonetheless, a factor of a few discrepancy in size scale and/or issues
with the detailed shape of the light curve is clearly less problematic
than the original factor of $\gtrsim 10^4$ difference in timescale. Hence  
some form of X-ray reprocessing is the favoured model for the stochastic, 
small scale (a few tens of percent) optical variability 
(see e.g., \citealt{2012MNRAS.423..451L}). However, this is much more
difficult for systems which show larger amplitude variability. Some of
these can be due to extrinsic effects, such as obscuration events,
where material from the clumpy molecular torus moves into or out of
the line of sight (e.g., \citealt{2012ApJ...747L..33E}; \citealt{2017MNRAS.470.3591G}), 
or microlensing (e.g., \citealt{2016MNRAS.463..296L}), but clearly there
is an intrinsic change in luminosity in systems where the
emission lines also respond. In particular, a small subset of AGNs
change from type 1 (showing both broad and narrow lines) to type 1.9
(where the broad lines almost disappear) or vice versa, associated
with a change by more than a factor 2 in continuum flux (``changing-look'' 
AGN: see e.g., \citealt{2015ApJ...800..144L} and references therein).
Some of these also show more extreme variability associated with the
``changing look'', where the continuum changes by 
more than a factor of $\gtrsim 10$ over a timescale of years (e.g.,
\citealt{2017MNRAS.470.4112G}). These objects are rare, but are
now being increasingly discovered in large surveys with repeat cadence
(e.g., \citealt{2016MNRAS.457..389M}).

We focus here on one particularly well studied case of extreme
variability associated with a ``changing look''. The bare (low
intrinsic absorption) nucleus Mrk~1018 changed from type 1 to 1.9
around 1980 \citep{1986ApJ...311..135C} and from type 1.9 to 1 around
2010 \citep{2016A&A...593L...8M}. This event is unlikely to be
associated with uncovering/covering of the central AGN by a dusty
cloud from the torus as the timescale does not correspond to the
expected orbital velocity out at torus size scales, nor is there any
progressive change in reddening as the source brightens/dims.
Instead, for an intrinsic change in UV continuum, the BLR clouds
should respond to the changing photo-ionizing flux (LOC:
\citealt{1995ApJ...455L.119B}; \citealt{1997ApJS..108..401K};
\citealt{2000ApJ...536..284K}), while the NLR stays constant due to
the much longer light travel timescales.

The broad lines are most sensitive to the extreme ultraviolet (EUV) 
part of the spectral energy distribution (SED), and this ionizing flux
is mostly unobservable due to galactic absorption. Nonetheless, 
multiple studies show that the continuum emission turns down below the
expected disc emission as the spectrum approaches the Lyman limit at
$\sim 1000$~\AA, while it turns up above the extrapolated 2--10~keV X-ray
power-law emission in the soft X-ray region. These two spectral breaks
point towards each other, and can be fitted by a Comptonization spectrum
dominating the ionizing flux in the FUV region. The parameters of this
Comptonization region are remarkably similar across multiple AGNs,
indicating that there is ubiquitous warm ($kT_{\rm e} \sim 0.2$~keV),
optically thick material ($\tau\sim 10$--$20$: e.g., \citealt{2003A&A...412..317C};
\citealt{2004MNRAS.349L...7G}; \citealt{2011A&A...534A..39M}; 
\citealt{2011PASJ...63S.925N}; \citealt{2013PASJ...65....4N}; \citealt{2014MNRAS.439.3016M}; 
 \citealt{2017arXiv171004940P}; \citealt{2018A&A...609A..42P}) as well as an inner hard X-ray corona
which is hot ($kT_{\rm e} \sim 50$--$100$~keV) and optically thin
($\tau \sim 1$). Thus to track the BLR ionizing flux, we need broad
band, simultaneous UV and X-ray datasets which can determine the warm
and hot Comptonization components as well as any standard disc
emission. 

The decline of Mrk~1018 from 2008 was fairly well monitored with 
both optical/UV and X-ray telescopes on \textit{Swift} and \textit{XMM-Newton} 
(\citealt{2016A&A...593L...8M}; \citealt{2016A&A...593L...9H}). 
Hence we can use these data to model the change in soft X-ray excess
and hard X-ray emission during the large drop in continuum as the 
source changed from type 1.0 to 1.9 over an 8 year period.

\begin{figure*}
	\includegraphics[width=160mm]{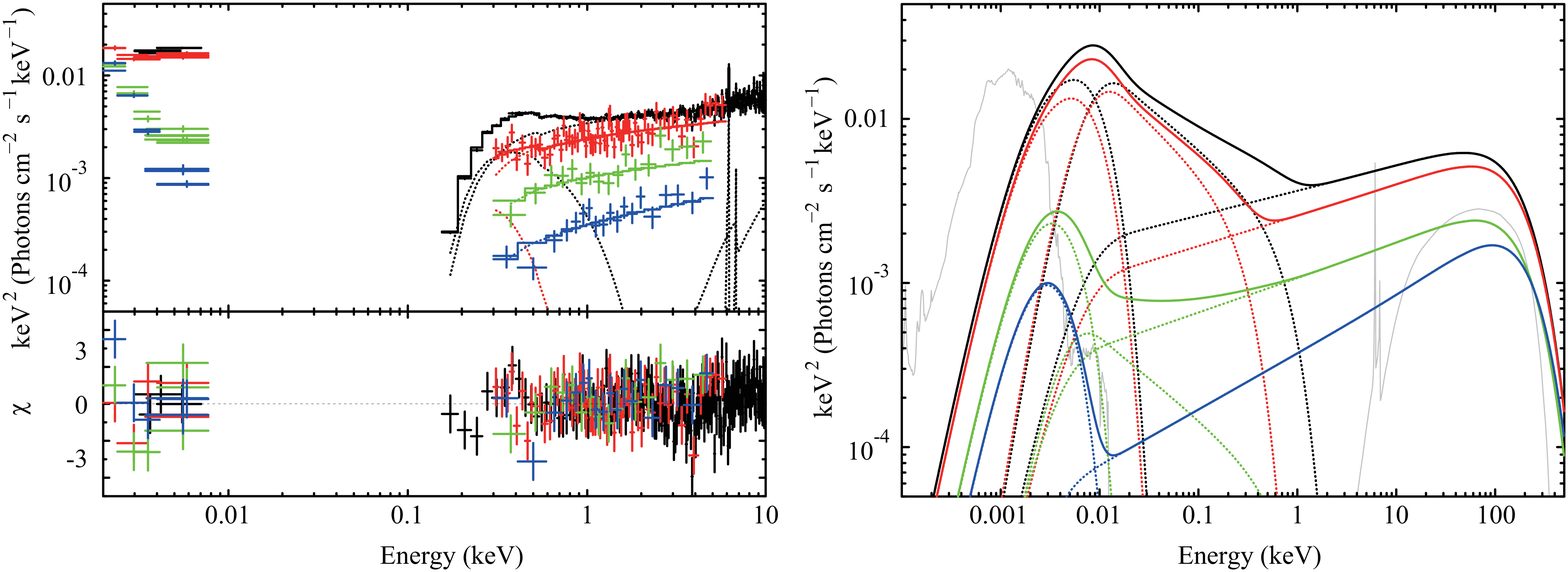}
    \caption{Left: Optical, UV, and X-ray spectra in 2008 Aug (black), 2008 Jun (red), 2013 (green), and 2016 (blue) in $\nu F_{\nu}$ forms, fitted with the model of \texttt{phabs*redden*(optxagnf + MYTorusS + MYTorusL + hostpol)}. Right: Best-fit AGN model spectra shown after the absorption and reddening are removed, and in the same colors as in left panels. Grey shows model spectra of an S0-type host galaxy emission, and a torus reflection component accompanied by fluorescence lines in 2008 Aug.}
    \label{fig:fig1}
\end{figure*}

\begin{figure*}
	\includegraphics[width=160mm]{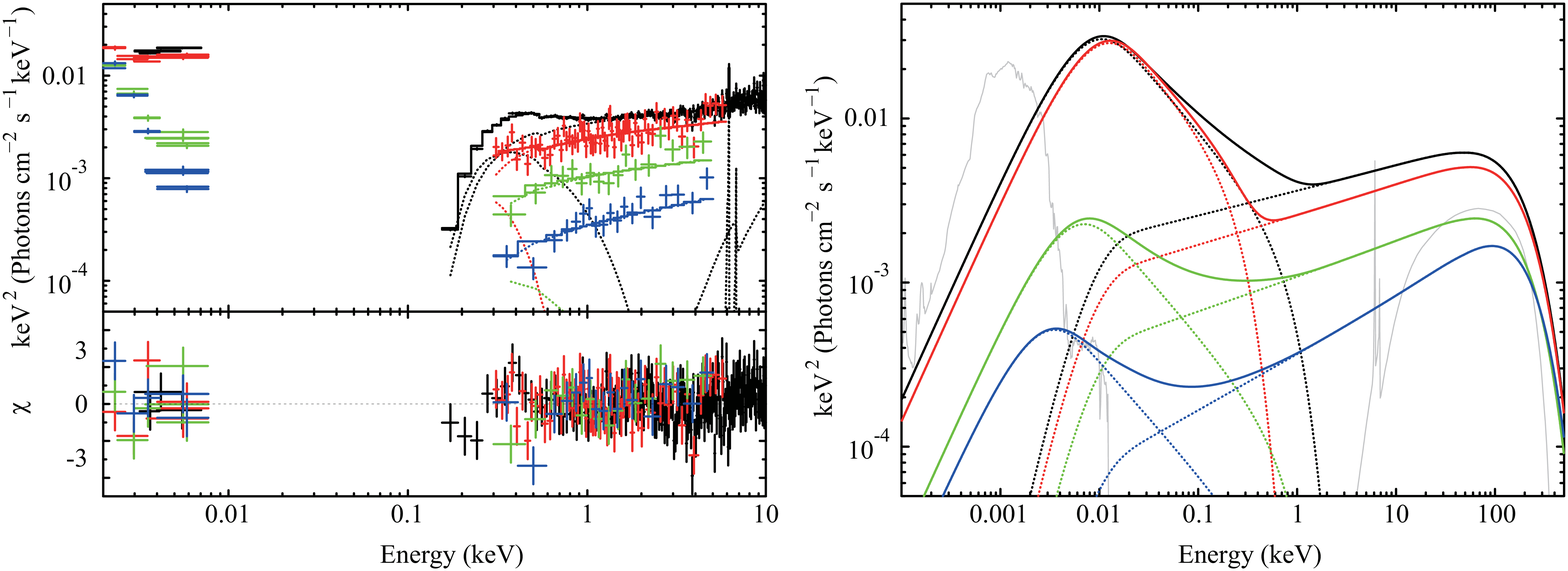}
    \caption{Same as Fig.~\ref{fig:fig1}, but with the model of \texttt{phabs*redden*(nthcomp\_S + nthcomp\_H + MYTorusS + MYTorusL + hostpol)}.}
    \label{fig:fig2}
\end{figure*}

\begin{table*}
 \caption{Model parameters obtained by the fits to the optical, UV and X-ray spectra in 2008 Aug, 2008 Jun, 2013, and 2016.}
 \label{tab:tab1}
 \begin{center}
  \begin{tabular}{ccccccc}
   \hline\hline
    Component & Parameter  & 2008 Aug & 2008 Jun & 2013 & 2016 \\
          \hline
                            	 
     \texttt{phabs} &$ N_{\rm H}~(10^{20}$~cm$^{-2}$)
			& \multicolumn{4}{c}{$2.43$~(fix)}\\[1.5ex]
     \texttt{redden} &$E(B-V)$
			&  \multicolumn{4}{c}{$= 1.5\times N_{\rm H}/(10^{22}$~cm$^{-2}$)}\\
      
      \hline
      
    \multicolumn{6}{c}{Fit with outer standard disc}\\[1.5ex]     	
                          
      \texttt{optxagnf} 
      			& $M_{\rm BH}~(M_{\odot})$ 
			& \multicolumn{4}{c}{$6.92\times10^7$~(fix)}\\
      
      			& $D$~(Mpc)
  			& \multicolumn{4}{c}{$176$~(fix)}\\
			
			& $\log (L/L_{\rm Edd})$
			& $-1.14^{+0.02}_{-0.03}$
			& $-1.26^{+0.04}_{-0.08}$
			& $-1.93^{+0.07}_{-0.06}$
			& $-2.27 \pm 0.03$\\

      			& $a^{*}$
  			& \multicolumn{4}{c}{$0$~(fix)}\\
			
			&$R_{\rm cor}~(R_{\rm g})$
			&\multicolumn{2}{c}{$60.0^{+71.8}_{-24.3}$}
			&\multicolumn{2}{c}{$73.7^{+52.0}_{-5.3}$}\\
			
			&$R_{\rm out}~(R_{\rm g})$
			& \multicolumn{4}{c}{= Self gravity radius $= 2.3\times10^3 (L/L_{\rm Edd})^{4/9}$} \\
			
			&$kT_{\rm e}$~(keV)
			& $0.17 \pm 0.02$
			& $0.07^{+0.07}_{-0.02}$
			&\multicolumn{2}{c}{$=kT_{\rm e}$ in 2008 Aug}\\
   				
			&$\tau$
			& $18.5^{+1.6}_{-1.2}$
			& $29.3^{+37.9}_{-27.3}$
			&\multicolumn{2}{c}{$=\tau$ in 2008 Aug}\\

			& $\Gamma_{\rm H}$
			& $1.85 \pm 0.03$
			& $1.82^{+0.07}_{-0.09}$
			& $1.79^{+0.10}_{-0.09}$
			& $1.66$~(fix)\\	
			
			& $f_{\rm PL} = L_{\rm H}/(L_{\rm S} + L_{\rm H})$
			& $0.48^{+0.22}_{-0.05}$
			& $0.47^{+0.10}_{-0.08}$
			& $0.91^{+0.06}_{-0.11}$
			& $0.85$--$1$\\	

      			& $z$
  			& \multicolumn{4}{c}{$0.0424$~(fix)}\\	
			
      			& $N$
  			& \multicolumn{4}{c}{$1$~(fix)}\\ [1.5ex]
			
   $\chi^{2}$/d.o.f. &  &  \multicolumn{4}{c}{326.5/267} \\					
\hline

    \multicolumn{6}{c}{Fit without outer standard disc}\\[1.5ex]

      \texttt{nthcomp\_S} 
      			& $\Gamma_{\rm S}$ 
			& \multicolumn{4}{c}{$2.68^{+0.15}_{-0.09}$}\\

			& $kT_{\rm e,S}$~(keV) 
			& $0.20^{+0.05}_{-0.02}$ 
			& $0.06^{+0.02}_{-0.01}$ 
			&\multicolumn{2}{c}{$=kT_{\rm e,S}$ in 2008 Aug}\\

			& $kT_{\rm bb,S}$~(eV) 
			& $3.01^{+0.54}_{-0.42}$ 
			& $3.62^{+1.10}_{-0.66}$ 
			& $1.99^{+1.00}_{-0.56}$ 
			& $1.00^{+0.61}_{-0.00\rm{b}}$\\	

      			& $z$
  			& \multicolumn{4}{c}{$0.042436$~(fix)}\\

			& $N_{\rm S}~(10^{-5})$ 
			& $48.11^{+10.29}_{-8.22}$ 
			& $(2.09^{+4.12}_{-0.98})\times 10^{-2}$ 
			& $2.71^{+1.63}_{-0.92}$ 
			& $0.39^{+0.21}_{-0.10}$ \\

			& $L_{\rm S}~(10^{44}~\textrm{erg/s})$ 
			& $6.30 \pm 0.16$ 
			& $5.56^{+0.20}_{-0.19}$ 
			& $0.48 \pm 0.04$ 
			& $0.11 \pm 0.02$ \\ [1.5ex]

      \texttt{nthcomp\_H} 

      			& $\Gamma_{\rm H}$
			& $1.85$~(fix)
			& $1.82$~(fix)
			& $1.79$~(fix) 
			& $1.66$~(fix) \\
			
			& $kT_{\rm e,H}$~(keV)
			&\multicolumn{4}{c}{$100$}\\

			& $kT_{\rm bb,H}$~(eV)
			&\multicolumn{4}{c}{$=kT_{\rm bb,S}$}&\\	

      			& $z$
  			&\multicolumn{4}{c}{$0.042436$~(fix)}&\\

			& $N_{\rm H}~(10^{-3})$ 
			& $3.63 \pm 0.05$ 
			& $2.60^{+0.11}_{-0.12}$ 
			& $1.09 \pm 0.11$ 
			& $0.37\pm 0.03$\\
			
		        & $L_{\rm H}~(10^{44}~\textrm{erg/s})$ 
			& $2.71 \pm 0.03$ 
			& $2.04 \pm 0.09$ 
			& $0.91 \pm 0.09$ 
			& $0.45 \pm 0.04$ \\[1.5ex]		

   $\chi^{2}$/d.o.f. &  &  \multicolumn{4}{c}{314.2/269} \\

\hline\hline

  \end{tabular}
\end{center}

\end{table*}


\section{Observations and Data Reduction}

We analyze the archival datasets of Mrk~1018 derived by the EPIC-PN
and OM onboard \textit{XMM-Newton}, and by the XRT and UVOT
onboard \textit{Swift}. We perform the following data reductions to the
individual satellite datasets.

\textit{XMM-Newton}: The source was observed on  2008 Aug 7
(obsid of 0554920301), hereafter 2008 Aug. 
We reprocess the PN and OM datasets 
using \texttt{epchain} and \texttt{omichain} in
\texttt{SAS-16.0.0}, respectively. We exclude PN events
when the 10--12~keV background count rate is more than
$0.35$~cnt~s$^{-1}$. The PN source spectrum is extracted from a
$40''$-radius circle region centered at the AGN, while the PN
background spectrum is from a $1'$-radius circle region off the
AGN. The PN response matrix and auxiliary response file are prepared
by the \texttt{rmfgen} and \texttt{arfgen} in \texttt{SAS-16.0.0},
respectively. The U, UVW1, and UVW2 spectra are created via
\texttt{om2pha} using count rates obtained by \texttt{omdetect} in the
sequence of \texttt{omichain} and the response files distributed by
the \textit{XMM-Newton} Science Operations Centre.

\textit{Swift}: The source was observed on 2008 Jun 11
(obsid: 00035776001), 2013 Mar 1 and Jun 7--8, 
(obsid: 00049654001 and 00049654002), and 
2016 Feb 11 and Feb 16 (obsid: 00080898001 
and 00080898002). 
To extract X-ray spectral files including background files, 
response matrixes and ancillary response files, we use the automated 
pipeline of the UK \textit{Swift} Science Data Center 
\citep{2007A&A...469..379E, 2009MNRAS.397.1177E}.
In the spectral extractions, we request to co-add the two datasets 
(00049654001, 00049654002) in 2013, 
and the two (00080898001, 00080898002) in 2016 to get 
better signal-to-noise, and select events with grade 0--12. 
These procedures provide three \textit{Swift}/XRT spectra, 
and hereafter we call them the X-ray spectra in 2008 Jun, 2013 and 2016. 
We extract the V, B, U, UVW1, UVM2, and the UVW2 count rates from 
datasets with obsid 00035776001, 00049654001 and 00080898001 
to make optical--UV spectra in 2008 Jun, 2013, and 2016, respectively. 
The commands of \texttt{uvotimsum} and
\texttt{uvot2pha} in \texttt{heasoft-6.20} are used to merge multiple
extensions, and to convert count rates into XSPEC readable files. 
The source region is a $5''$-radius circle centered at the AGN while the
background is taken from a $50''$-radius circle off the AGN, which
includes no other bright sources.  We use the response files of
the UVOT bands distributed by the UK \textit{Swift} Science Data
Center.

By the \texttt{grppha} command in \texttt{heasoft-6.20}, 
each X-ray spectrum is binned so that a bin includes at least 20~photons 
to appropriately employ $\chi^2$ statistics in spectral fits. 
After the binning, an energy range up to $\sim10$~keV is noticed 
in the \textit{XMM-Newton}/EPIC-PN spectrum, while we ignore 
the \textit{Swift}/XRT spectra in an energy range of $\gtrsim6$~keV 
where the number of photons is less than 20.
Unless otherwise stated, errors shown in tables and figures refer to
$90$\% and $1~\sigma$ errors, respectively.

\section{Data Analysis}

We first examine how the broadband spectrum of Mrk~1018 varies over
time.  Figure~\ref{fig:fig1} shows the spectra from optical to X-ray
in 2008 Jun (red) and Aug (black), 2013 (green), and 2016 (blue). It
is clear that there is considerable difference in the variability
across the spectrum. The optical remains fairly constant, indicating a
large contribution from the host galaxy, while the UV drops by over an
order of magnitude. The X-rays also dim, but by a smaller factor, and
change shape from that showing a strong soft X-ray excess below 1~keV 
in the brightest spectrum (2008 Aug: black) which disappears in the dimmer
spectra.

\subsection{Outer disc, soft Comptonization, hard Comptonization model}

To quantify the spectral shape variations in optical, UV and X-ray, we
first perform the broadband AGN spectral fit with a standard SMBH
accretion model containing the disc emission, the soft X-ray excess,
and the hard Comptonization continuum \texttt{optxagnf}
\citep{2012MNRAS.420.1848D}, modified by the Galactic photo-absorption
and reddening effects modeled by \texttt{phabs} and \texttt{redden},
respectively. This assumes a standard disc emissivity
(Novikov-Thorne), but that this energy is only dissipated in a
standard disc structure from the outer radius 
$R_{\rm out}$ to a coronal radius $R_{\rm cor}$ which is a free
parameter. Within $R_{\rm cor}$, the gravitational energy is split, with a
fraction $f_{\rm PL}$ powering the coronal X-ray power-law emission,
characterized by a photon index $\Gamma_{\rm H}$, and the remainder dissipated
in a warm Comptonization component, characterized by electron
temperature $kT_{\rm e}$ and optical depth $\tau$.  This model has enough
flexibility to fit to AGN spectra, but with the physical constraint
that all the emission is assumed to be powered by a constant mass
accretion rate, parameterized as $L/L_{\rm Edd}$ for a given SMBH
mass $M_{\rm BH}$ and spin $a^{*}$. This additional energy conservation
assumption allows the model components to be constrained across
the unobservable EUV region. 
The model then is normalized by the physical parameters of $M_{\rm BH}$, 
$L/L_{\rm Edd}$, $a^{*}$ and co-moving distance $D$.  Mrk~1018 has single
epoch black hole mass of $\log (M_{\rm BH}/M_\odot) = 7.4$--$7.9$ \citep{2016A&A...593L...8M}. 
We use $\log (M_{\rm BH}/M_\odot) = 7.84$ in all that follows \cite{2017MNRAS.472.3492E}, 
and assume $a^{*}=0$
following previous SED fits to the 2008 data with {\tt optxagnf}. 
We use the co-moving distance of $D \sim 176$~Mpc, corresponding to 
the source redshift of $z = 0.042$.

The X-ray data also show a clear Fe-K$\alpha$ at $\sim 6.4$~keV in the
\textit{XMM-Newton} spectrum of 2008 Aug. The remaining \textit{Swift} datasets have
insufficient signal-to-noise to constrain this feature, but comparison
of a \textit{Chandra} observations from 2010 and 2016 (not included here as
these have no simultaneous optical/UV data) shows that the line is
narrow, and neutral, consistent with an origin in the torus \citep{2017ApJ...840...11L}. 
Hence, from physically-motivated torus reflection models called \texttt{MYTorus}, 
we include a Compton-scattered component and emission lines 
(hereafter we call \texttt{MYTorusS} and \texttt{MYTorusL} respectively),  
calculated with a Comptonized thermal intrinsic continuum with the 
electron temperature of $100$~keV.  
We fix the optical depth for the intrinsic continuum, inclination angle, and the 
redshift at 0.8, 0$^\circ$, and 0.0424, respectively, while the column density of the torus
and the normalization are tied between \texttt{MYTorusS} and \texttt{MYTorusL}, and left free. 
In 2008 Jun, 2013, and 2016, the torus reflection cannot be well constrained due to
the usable energy range (see \S2), we set the normalizations in \texttt{MYTorusS} and 
\texttt{MYTorusE} to zero. 

The optical data clearly also show a constant host galaxy
component. This is a late type merger, categorized as a lenticular
galaxy (the S0 type in the Hubble classification, e.g. \citealt{2010ApJ...710..503W}). 
Following \cite{2017MNRAS.472.3492E}, we add a model of S0-type host galaxy
emission template named \texttt{hostpol} \citep{2007ApJ...663...81P}.  
We tie the normalization of \texttt{hostpol} across the four spectra.
The total model fit to the four datasets simultaneously is then 
\texttt{phabs*redden*(optxagnf + MYTorusS + MYTorusL + hostpol)}.
The column density $N_{\rm H}$ of \texttt{phabs} is fixed at the
Galactic value of $2.43\times10^{22}$~cm$^{-2}$, while $E(B-V)$ of
\texttt{redden} is tied to $1.5 \times N_{\rm H}/(10^{22}$~cm$^{-2}$) utilizing 
the Galactic gas-dust ratio.
The \textit{Swift} data especially cannot constrain all the parameters of the
{\tt optxagnf} model. We tie the coronal outer radius $R_{\rm cor}$ 
between 2008 Jun and Aug, and between 2013 and 2016. 
In addition, we fix $\Gamma_{\rm H}$ in 2016 at 1.66 
following the 2016 phenomenological fit result without the \texttt{MYTorus} 
models in \cite{2017ApJ...840...11L}, considering that the normalizations 
of the \texttt{MYTorus} models are fixed at zero.

We first fit the four spectra by making the electron temperature
$kT_{\rm e}$ and optical depth $\tau$ of the soft Comptonization in
\texttt{optxagnf} free and tied among all. In the fit, the systematic
errors of 3\% are employed to account for the calibration
uncertainties mainly in absolute flux scaling between the OM and UVOT
(e.g., \citealt{2016MNRAS.457...38L}). The fit is not acceptable, with
$\chi^2$/d.o.f.$=347.5/269$, mainly because of large negative
residuals in the soft X-ray band in 2008 Jun, showing that the soft
X-ray excess spectral shape is different between 2008 Jun and
Aug. Allowing $kT_{\rm e}$ and $\tau$ in 2008 Jun to differ from those
in 2008 Aug gives an almost acceptable fit with $\chi^2$/d.o.f.$=326.5/267$.
Figure~\ref{fig:fig1} left shows the data fit to this model, together with
the residuals, while Fig.~\ref{fig:fig1} right shows the unabsorbed model
components. The best fit parameter values and 90\% confidence
uncertainties are given in Table~\ref{tab:tab1}.

\subsection{No outer disc model}

Recently, the optical/UV emission from an AGN has been interpreted by
only a soft X-ray excess component without an outer standard 
disc (e.g., \citealt{2015A&A...575A..22M}; \citealt{2017arXiv171004940P}). 
However, there is still implicitly an outer disc as the model requires 
that there is some source of seed photons for the warm Comptonization. 
\cite{2017arXiv171004940P} assume that these seed photons have 
a disc blackbody distribution, and this is a key to their optical/UV slope 
being able to fit many of their objects. The maximum seed photon 
temperature of the disc blackbody, $kT_{\rm bb,S}$, is left as a 
free parameter. This contrasts with the {\tt  optxagnf} model where 
the original picture was that the seed photons came from the disc 
underlying the warm Comptonization region (see Appendix A in 
\citealt{2012MNRAS.420.1848D}), but since these would affect the spectrum only 
in the unobservable EUV regime the code approximated these 
by blackbody seed photons at the temperature of the hottest part
of the standard disc i.e. $T(R_{\rm cor})$.
Thus we try the \cite{2017arXiv171004940P} model of 
\texttt{phabs*redden*(nthcomp\_S + nthcomp\_H + MYTorusS + MYTorusL + hostpol)}. 
The warm Comptonization is modeled by 
\texttt{nthcomp\_S}, while \texttt{nthcomp\_H} models the coronal
power-law emission, and both are assumed to have the same seed photon 
distribution, characterized by a disc blackbody with maximum 
temperature $kT_{\rm bb,S}$.

Similarly to the {\tt optxagnf} fits above, the data strongly require
that the soft X-ray excess shape in 2008 Jun is different from 2008 Aug. 
Hence if we tie the photon index of the warm Comptonization
between the four spectra then the data strongly require that its
electron temperature is different. Conversely, 
if we tie the electron temperature then we require that 
the photon index of the warm Comptonization changes. 
We choose then to tie the warm Comptonization
spectral index $\Gamma_{\rm S}$ between the four spectra, as
\cite{2017arXiv171004940P} give some physical reasons for fixing this from
reprocessing of the warm Comptonization in a passive, optically thick
disc beneath the corona. Because the warm Comptonization component is
not significantly detected in the soft X-ray emission in 2013 and
2016, we tie the \texttt{nthcomp\_S} coronal temperature $kT_{\rm e,S}$ 
across all the 2008 Jun, 2013, and 2016 datasets. 
We allow the seed photon temperature to be free in all datasets. 
The coronal Comptonization seed photon temperature is set to that of
the soft Comptonization in all datasets, but it is assumed to be
blackbody rather than disc blackbody. The photon index $\Gamma_{\rm H}$ 
is fixed at those obtained by the fit with \texttt{optxagnf}
to compare the reproducibility of the optical--UV band between the two
models, and we fix the \texttt{nthcomp\_H} coronal temperature
$kT_{\rm e,H}= 100$~keV as in {\tt optxagnf}. 
Reflection, host galaxy and absorption are included as before. 

This somewhat different spectral decomposition gives a similarly good fit with 
$\chi^2$/d.o.f.$=314.2/269$ (see table~\ref{tab:tab1}), showing that the 
optical can be modeled either by assuming an outer standard disc or 
by assuming that there is optically thick material which reprocesses 
the emission from a warm Comptonizing layer above it. The best fit model 
is shown against the data in Fig.~\ref{fig:fig2} left and as unabsorbed model 
components in Fig.~\ref{fig:fig2} right.

\subsection{Changing broadband continuum spectral shape}

\begin{figure}
	\includegraphics[width=80mm]{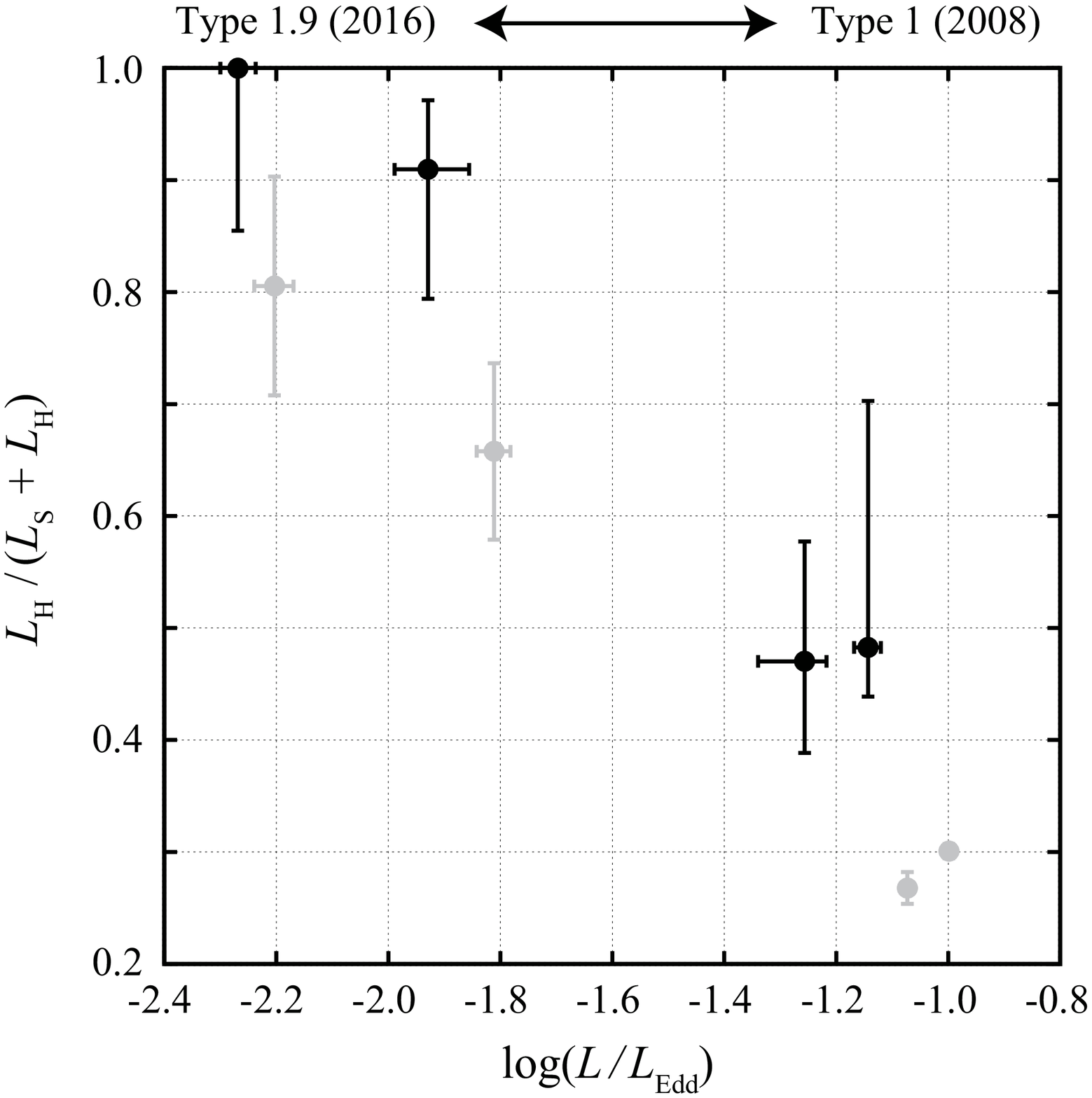}
    \caption{The $L/L_{\rm Edd}$-dependence of the fraction of the soft and hard Comptonization strengths written as $L_{\rm S}$ and $L_{\rm H}$ respectively. 
    These are obtained by the spectral fits with \texttt{optxagnf} (black) and \texttt{nthcomp\_S} + \texttt{nthcomp\_H} (grey) summarized in Fig.~\ref{fig:fig1}, Fig.~\ref{fig:fig2}, and Table~\ref{tab:tab1}. Errors here refer to 90\% errors.}
    \label{fig:fig3}
\end{figure}

These models are both acceptable fits to the data, and 
allow us to characterize how the accretion flow emission changes as a
function of $L/L_{\rm Edd}$. This is a direct parameter of the {\tt optxagnf} 
modeling, but is not explicitly included in the warm and hot Comptonization 
description. Instead, we integrate the model components to get 
the luminosity in the warm and hard X-ray Comptonizing regions. 
This allows us to calculate $L/L_{\rm Edd}$ using the same mass 
and distance as before. 

Both models have hard and soft Comptonization components, so
Fig.~\ref{fig:fig3} shows the change of $L_{\rm H}/(L_{\rm S} +
L_{\rm H})$, which is equivalent to $f_{\rm PL}$, when we use 
\texttt{optxagnf} (black) and the two Comptonization components 
with no outer disc (grey).  Both models show similar behaviour in 
that the warm and hot Comptonization components show
comparable luminosity at $L/L_{\rm Edd}\sim 0.08$, but that the ratio of the
hot component gets larger when $L/L_{\rm Edd}$ becomes smaller, and it
dominates the spectrum at $L/L_{\rm Edd} \sim 0.006$.
Note that the errors in the case of the two Comptonization components 
are relatively small mainly because the spectral shapes are fixed.

\subsection{Summary of results}

\begin{enumerate}

\item The observed 2--10~keV flux drops by a factor of 
 $6$--$7$ over $\sim 8$~years, while the UV luminosity 
 drops by a factor of $\sim 20$. 

\item Broadband spectral models show that the  
total luminosity drops by a factor of $13$--$15$, 
from  $L/L_{\rm Edd} \sim 0.08$ to $\sim 0.006$. 

\item The warm Comptonization region (soft X-ray excess) is very
  obvious when the source is brightest, where the broadband SED looks
  like a typical Seyfert type 1.

\item The warm Comptonization region may disappear completely as the
  source dims if there is residual emission from an outer standard
  disc ({\tt optxagnf} model). Alternatively, the faint optical/UV
  emission may be produced entirely by the warm Comptonization, and
  not require any outer standard disc, in which case the warm
  Comptonization drops by a factor of $\sim 60$.

\end{enumerate}

Previous models of these data assumed
a simple disc and power law model for the optical/UV and X-ray flux,
respectively, and did not specifically fit to the soft X-ray excess
\citep{2016A&A...593L...9H}. This is important as Fig.~\ref{fig:fig1}
and \ref{fig:fig2} show that almost all of the EUV photons, the ones which
most contribute to the BLR ionization, are in the soft excess
emission.  Therefore, the changing-look phenomenon in Mrk~1018 from
type 1.9 to 1 around 1980 \citep{1986ApJ...311..135C} and from type 1
to 1.9 around 2010 \citep{2016A&A...593L...8M} is clearly controlled
by the increase and decrease of the warm Comptonization region,
respectively. We will explicitly explore the effect of this 
changing spectral shape as well as luminosity on the predicted BLR
emission from an LOC model in a following paper.

The soft X-ray excess probably forms the seed photons for the 
hot Comptonization component, so a decrease in the ratio between 
soft and hard Comptonization means a decrease in seed photons for 
Compton cooling of the hard X-ray emitting electrons relative to the 
power injected into heating these electrons (e.g., \citealt{1999ASPC..161..295B}).
Hence, the large decrease of the warm Comptonization flux is likely to 
lead to spectral hardening of the hot Comptonization continuum, 
although significant photon index decrease of the 
hot Comptonization component is not supported by the data. 
This can be tested by longer X-ray observations especially of 
faint epochs in the future. 

However, the remaining results are much more difficult to explain, 
as discussed below.

\section{Discussion}

\subsection{Standard accretion disc timescales}
\label{s4-1}

Fundamentally, we see a drop in total accretion luminosity by a factor
of $13$--$15$ over a timescale of $\sim 8$ years. This is a problem
for any standard disc model.  There is a hierarchy of timescales for a
standard thin disc, with the dynamical timescale being fastest, then
the thermal timescale and then the viscous. We calculate these for
Mrk~1018, assuming that the optical/UV (typical temperature $3\times
10^4$~K) is emitted from a standard thin disc at $R \sim 100~R_{\rm g}$. 
This implies $H/R=c_s/v_\phi \sim 5\times 10^{-4}$ (where the sound speed
$c_s=\sqrt{kT/m_p}$ and Keplarian velocity
$v_\phi=\sqrt{GM/R}=\sqrt{R_{\rm g}/R}c$). The optical emission regions in
AGN discs are much thinner than the optical parts of cataclysmic variables (CV) or BHB discs,
as in AGN the optical is produced at $\sim 100~R_{\rm g}$ rather than
$10^5~R_{\rm g}$ so while the gas temperature is the same for the optical
emission, the Keplarian velocity on the denominator is much higher.
For $\alpha=0.1$ these give expected timescales of

\begin{eqnarray}
 t_{\rm dy} &\approx&  \left(\frac{GM_{\rm BH}}{R^3}\right)^{-\frac{1}{2}} \approx 4~\textrm{days} \\
 t_{\rm th} &\approx& \frac{t_{\rm dyn}}{\alpha} \approx 1~\textrm{month} \\
 t_{\rm vis} &\approx& \frac{t_{\rm dyn}}{\alpha}
 \left(\frac{H}{R}\right)^{-2}  
\approx 4\times 10^5~\textrm{years}.~~~
\end{eqnarray}

The observed timescale of $\sim 8$~years is much shorter than the
viscous timescale, while somewhat longer than the thermal and
dynamical timescales. The thermal timescale in particular
is suggested as the origin of the $10$--$20$\% stochastic optical variability which
is typically seen in AGNs on 
timescales of months (e.g.,
\citealt{2009ApJ...698..895K}; \citealt{2010ApJ...721.1014M}). 
The magnetorotational
instability (MRI) \citep{1991ApJ...376..214B}
produces fluctuations in dissipation, 
leading to changes in local heating on a thermal timescale
(e.g., \citealt{2011ApJ...727L..24D}).

However, this stochastic variability on timescales of months is 
very different to the factor of at least $\sim 20$ systematic decrease in
optical/UV over a period of $\sim 8$~years 
seen here. The amplitude of bolometric change requires that there is a 
change in either mass accretion rate or efficiency (or
both) of the accretion flow, yet this change occurs on 
a timescale which is too fast for the
mass accretion rate to change through a standard disc (viscous
timescale) by many orders of magnitude. 

\subsection{Disc instability in BHBs}

Variability in BHBs is much easier to study directly as the timescales
are much shorter, of order days-months-years.  There is now a good
general understanding that their transient outbursts, where the mass
accretion rate increases by orders of magnitude from quiescence to
outburst, are triggered by the Hydrogen ionization (H-ionization) disc
instability mechanism (DIM) in the outer disc. Mass accreting from the
companion star accumulates at the circularization radius.  This is
initially cold, below the H-ionization point, so there are no free
electrons and the opacity is low. The mass accretion rate through this
structure is lower than the rate at which mass is added from the
companion star so the surface density and temperature build up
together until it reaches a temperature of $\sim 3000$~K, at which
point part of the blackbody flux can start to ionize Hydrogen. This
traps these photons in the material, so its temperature rises, which
means more photons can ionize Hydrogen. This continues until all of
the Hydrogen is ionized, i.e to a temperature of $10^4$~K. This jump
in temperature also gives a jump in sound speed, so matter can be
transported through the disc faster i.e., the local accretion rate can
now be larger. This can trigger the next ring inwards to go unstable
to H-ionization, turning the local instability into a global switch in
disc properties if the difference in mass accretion rate is large
enough.  This requires a change in the effective viscosity, often
parameterized as a jump in $\alpha$  from neutral to
ionized, perhaps as an outcome of the MRI either requiring free electrons
\citep{1999ApJ...520..276M} or driving turbulence \citep{2016MNRAS.462.3710C}.  This
results in a heating front propagating inwards, traveling at a speed
which is shorter than the viscous timescale by a factor $(H/R)$
 (\citealt{1998MNRAS.298.1048H};
\citealt{2001A&A...373..251D}; \citealt{2001NewAR..45..449L}; \citealt{2009A&A...496..413H}).

The mass accretion rate through the disc is now larger than the mass
accretion rate from the companion star, so it cannot maintain this
steady state structure. When the outer disc temperature falls below
$10^4$~K, the instability works in reverse, triggering a cooling wave
through the disc which switches Hydrogen back to neutral and
propagates at a speed faster than the heating front by a factor
$\alpha_{\rm h}/\alpha_{\rm c}$.  

In accreting white dwarf systems, the cooling wave is triggered
quite quickly after the heating wave, shutting off the outburst after 
only a small amount of the total disc material is accreted \citep{1996PASP..108...39O}. 
Instead, in BHBs and neutron stars, the copious X-ray emission
from the inner disc which is associated with the much larger
gravitational potential irradiates the outer disc material, keeping it
ionized. The disc material accretes viscously (exponential decline)
until the X-ray luminosity drops enough for the outer disc to dip
below the H-ionization temperature. The cooling front starts, but its
inward speed is set instead by the viscous shrinking of the irradiated
zone, giving a linear decline \citep{1998MNRAS.293L..42K}.

Thus the mass accretion rate onto the black hole can increase by many
orders of magnitude on timescales of days for the heating front
propagation \citep{2001A&A...373..251D}, while the decline is rather slow as it
is controlled by the viscous time of the outer disc. 
This mechanism is well tested in BHBs and neutron stars 
(\citealt{1999MNRAS.306...89S}; \citealt{2012MNRAS.424.1991C}), 
and in white dwarfs \citep{1996PASP..108...39O}. 

\subsection{State change in BHBs}

The dramatic variability associated with the H-ionization DIM can
also trigger a spectral transition, which most probably represents a
switch in the nature of the accretion flow from a hot, optically thin,
geometrically thick flow (low/hard, Comptonization dominated e.g. the
advection dominated accretion flow: \citealt{1995ApJ...452..710N}) to a cool,
optically thick, geometrically thin disc (high/soft, thermal
dominated). This transition can occur over a wide range in luminosity
on the fast rise to outburst, but this may be due to the flow being
far from steady state (\citealt{2002ApJ...569..362S};  \citealt{2007MNRAS.378...13G}). 
Instead, the reverse transition on the slow decline occurs at a
rather stable luminosity around $0.02~L_{\rm Edd}$ (\citealt{2003A&A...409..697M},
see e.g. the review by \citealt{2007A&ARv..15....1D}).

Cygnus X-1 shows this spectral transition decoupled from the DIM as its
outer disc is stable. This shows that the hard state is on average a
factor of $\sim3$--$4$ lower in bolometric luminosity than the soft
state \citep{2002ApJ...578..357Z}. This could be entirely produced by
the state change alone, with a constant mass accretion rate as the
hard state should be somewhat radiatively inefficient, and some of the
accretion energy is required to power the potentially substantial
kinetic luminosity of the compact jet seen in the radio emission
\citep{2002ApJ...578..357Z}.  However, some change in mass accretion
rate is required to trigger the state change, so it seems more
reasonable to assume that there are long timescale fluctuations in
mass accretion rate of perhaps a factor of $\sim 1.5$--$2$ so that the
bolometric luminosity at a single transition only changes by a factor
$\sim 1.5$--$2$. Thus a state change can give a drop in luminosity by
a factor $2$--$4$ {\em without} requiring any change in mass accretion
rate. 

The soft to hard transition involves a quite dramatic decrease in
surface density and concomitant increase in temperature (see e.g.  the
review by \citealt{2014ARA&A..52..529Y} Fig.~3). Changing the 
surface density radially requires a viscous timescale but here
the disc evaporates 
so it can lose mass vertically as it heats up to the virial temperature.
\cite{1986A&A...159L...5S} show that even a standard Shakura-Sunyaev
disc will have a thermal instability in its outer layers, as by
definition the photosphere is rather inefficient at cooling, so any
heat dissipated here heats up this material to the virial
temperature. The next layer of the disc is then the photosphere, so it
too heats up, and this process can lead to the entire disc switching
into the hot solution (see e.g. \citealt{2013MNRAS.435.2431D}; \citealt{2017ApJ...843...80H}).
However, while this does involve vertical rather than radial motion, the timescales are still the
radial viscous timescale \citep{2006ASPC..360..265C} as can be seen most clearly from the 
reverse process of condensation at the soft-to-hard transition. The hot flow collapses down
into the midplane, but its surface density is so much smaller than that of a standard disc that it
requires a viscous timescale for radial accretion from the outer disc to build up the 
material back into a standard disc. Indeed, this is seen in BHBs, where the timescale for the 
soft-to-hard 
state transition is $\sim 0.5$~days which is
close to the thin disc viscous timescale at a few hundreds $R_{\rm g}$
which is the expected radius of the evaporated region \citep{2007MNRAS.376..435M}.

\subsection{Differences between AGNs and BHBs}

Fundamentally, the behaviour of Mrk~1018 looks like a large drop in luminosity, similar to that seen in 
the transient BHBs which are triggered by the DIM, together with a state change as the luminosity drops below
a few percent of Eddington. However, as has often been noted, the timescales 
predicted by the thin disc models or by scaling from galactic binaries are far too long. Here we 
critically assess these timescales, pointing out some significant differences between AGNs and BHBs. 

A key difference between AGNs and BHBs is that the disc in AGNs is
self-gravitating at around a few hundreds $R_{\rm g}$ (e.g.,
\citealt{1989MNRAS.238..897L}). This outer disc radius is much smaller
than the disc in BHBs, which is truncated by tidal effects from the
companion star at typical radii of $R\sim 10^5~R_{\rm g}$.  The
combined effect of mass and radius predicts all timescales scale as 
$M r_{\rm out}^{3/2}$ where $R_{\rm out}=r_{\rm out}R_{\rm g}$.

Another key difference, connected to the smaller $r_{\rm out}$ in AGNs, is
in the importance of irradiation. This is effective only at large
radii (\citealt{1975ApJ...202..788C}; \citealt{2016ApJ...823..159S}) so becomes mostly
unimportant in AGN discs \citep{2017MNRAS.470.3591G}.  Thus the DIM in AGNs
should be more comparable to the DIM in white dwarf systems, where
irradiation is not important due to the much weaker white dwarf
gravitational potential \citep{2009A&A...496..413H}. Thus the rise and
decay times in AGNs should be set by the heating/cooling front
propagation speeds, not by the viscous decay timescale.

We scale the DIM timescale from the observed dwarf novae
$\sim1$~day timescales of the heating/cooling front for a $1~M_\odot$
white dwarf with $r_{\rm out}\sim 10^5$ up to an AGN with mass $7\times
10^7~M_\odot$ and   $r_{\rm out} \sim 10^2$. This predicts 
a timescale of $\sim 6$~years for the front propagation in AGNs, similar to
that observed. 

However, this assumed that $H/R$ is similar in the AGN disc as in
CV's, whereas a standard thin disc has $H/R=c_{\rm s}/v_\phi\sim
(kT_{\rm gas}/m_{\rm p}c^2)^{1/2} (1/r_{\rm opt})^{1/2} $ which is much smaller in
AGNs than in binaries (see \S4.2 above).  The front propagation timescale
depends on $R/H$ i.e. is a factor 30 longer than a simple scaling from
CV's, lengthening the predicted timescale to $\sim1000$~years. This is comparable to 
the timescales seen in the AGN DIM simulations of \cite{2009A&A...496..413H}.
Nonetheless, the data clearly show that the electron temperature in the soft X-ray excess region is
two orders of magnitude higher than expected from the thin disc equations
(0.2~keV rather than 2~eV). If this sets the scale height then AGN discs are more 
comparable in $H/R$ to the CV discs, but it is very unclear how the DIM would then operate
as the electron temperature is far above the H-ionization temperature required for the 
heating/cooling wave trigger.

There is one more break in the scaling between AGN and binaries, which
is the importance of radiation pressure. The lower densities typical of AGN
discs mean that gas pressure is lower at the same temperature, so
radiation pressure is more likely to dominate. Thus the sound speed is
set by the radiation pressure rather than the gas pressure, and is
faster than in gas pressure dominated discs. For Mrk~1018 at its peak luminosity, 
radiation pressure is a factor $\sim 75$ larger than the gas pressure in the optical emission 
region (e.g., \citealt{1989MNRAS.238..897L}). 
Radiation pressure was not included in the AGN DIM simulations of \cite{2009A&A...496..413H}
as radiation pressure dominated discs are unstable \citep{1973A&A....24..337S}. This radiation pressure
thermal-viscous instability is seen in all simulations where heating scales with total (gas plus radiation) pressure
(e.g., \citealt{2001MNRAS.328...36S}), resulting in characteristic limit cycles between a slim, super-Eddington inner disc and a much dimmer standard disc at all luminosities $L>0.05~L_{\rm Edd}$. These are not typically observed in BHBs
in this luminosity range (e.g., \citealt{2004MNRAS.349L...7G}), which could indicate that they are stabilized 
by magnetic pressure (e.g. \citealt{2003A&A...412..317C}; \citealt{2017A&A...603A.110G}). The total pressure should be even larger than that predicted by the radiation pressure dominated disc, giving an even faster sound speed for the heating/cooling front, though we note that current MRI simulations do not show magnetic pressure dominating
as they show the instability \citep{2013ApJ...778...65J}.

The large drop in luminosity could then be the DIM cooling front, propagating through 
a radiation and/or magnetic pressure dominated AGN disc. As the luminosity drops below a few percent of Eddington, it triggers the state transition as seen in BHBs. This occurs at more or less constant luminosity, but still requires a 
viscous timescale to change the surface density by evaporation/condensation. Scaling the BHBs
soft to hard transition timescale of $\sim 0.5$~days by the mass difference alone (as both only affect 
the inner $100~R_{\rm g}$ gives $\sim 10^4$~years, but this does not take into account the difference in $H/R$ 
between the two systems, making the predicted timescale even longer. 
Again, this can be solved if the disc is dominated by radiation and/or 
magnetic pressure instead of  gas pressure. 
Once the system has made the transition to a hot accretion flow it should vary on the viscous timescale of the hot flow, which is  of order the thermal timescale as $H/R$ is of order unity.

\subsection{Observational predictions}

As we suggest in \S4.4, Mrk~1018 likely experiences both drop in mass accretion rate 
due to the cooling front propagation and spectral state transition by disc evaporation. 
Considering these processes individually, extremely variable AGNs are divided into 
three groups; (1) sources showing  a state change due to disc evaporation/condensation 
associated with a factor 2--4 decrease/increase in luminosity, 
(2) sources with large mass accretion rate change due to the 
thermal front propagation due to the H-ionization instability, and (3) sources showing both. 

We suggest that many of changing-look AGNs are included in the group (1) or (3), 
and their variations commonly cross the state transition boundary of 
$L/L_{\rm} \sim$ a few percent.  
Sources with the variability amplitude of a factor a few are probably categorized into 
group (1), while those with amplitude of factor more than 10 should be into group (3). 
Interestingly, many of the known changing-look AGNs, e.g., NGC~2617
\citep{2014ApJ...788...48S}, Mrk~590 \citep{2014ApJ...796..134D}, and
SDSS~J0519+0033 \citep{2015ApJ...800..144L} actually show $L/L_{\rm Edd} \sim$
a few percent. 
A similar drop in the soft X-ray excess is seen by \cite{2012ApJ...759...63R} 
in Mrk~590 between 2004 and 2011, which we suggest is due to a state change 
associated with the changing-look phenomena, as in Mrk~1018. 
Recently, several tens changing-look AGNs have been 
newly reported (\citealt{2016MNRAS.457..389M}; \citealt{2017arXiv171108122Y}), 
but their Eddington ratios have not been reported yet. We thus predict 
that these new changing-look AGNs have $L/L_{\rm Edd} \sim$
a few percent as well.  Vice versa, we also predict that AGNs which 
exhibit drastic $L/L_{\rm Edd}$ variation crossing a few percent, e.g., 
NGC~3227 \citep{2014ApJ...794....2N} and
NGC~3516 \citep{2013ApJ...771..100N} will experience changing-look
phenomena simultaneously to drastic soft excess variation at
$L/L_{\rm Edd} \sim$ a few percent.  
We need further continuous optical spectroscopic monitoring  
spontaneously to soft X-ray monitoring on these sources for 
unconfirmed changing-look phenomena. 

As \cite{2010ApJ...721.1014M} reported, quasar variability cannot be explained by 
stochastic optical fluctuations driven by accretion rate, and there is additional 
variability depending on black hole mass and/or luminosity. 
If the thermal front propagation occurs as group (2), variability amplitude 
is determined by swept disc area which varies with black hole mass, i.e., 
luminosity with fixed accretion rate. Thus, the quasar variability may include 
that by the front propagation through a thin disc (e.g., \citealt{2009A&A...496..413H}), 
and some quasars are perhaps included into group (2) although 
they do not show the changing-look features. 
This can be tested by investigating detailed radial profile of disc temperature 
in quasars, which might include discontinuous drop if a thermal front is present. 
If their $L/L_{\rm Edd}$ could cross a few percent by the front propagation 
like Mrk~1018, they would show state transition, and enter group (3) being 
a changing-look quasar as reported by \citep{2015ApJ...800..144L}. 

\section{Conclusions}

In the present paper, we show for the first time how the changing-look phenomenon
is linked to a large change in the warm Comptonization component which forms
the soft X-ray excess.
The bolometric luminosity drops from $\sim 0.08~L_{\rm Edd}$ to $\sim 0.006~L_{\rm Edd}$,
but the hard X-rays drop by only a factor $\sim 7$ while the warm Comptonization
decreases by at least a factor $\sim 60$ and may disappear completely.
This marked hardening of the spectrum looks very like the transition
seen in stellar mass BHBs around $0.02~L_{\rm Edd}$,
which is generally interpreted as the inner regions of an optically
thick accretion disc evaporating into a hot accretion flow. We
suggest that all the true changing-look AGNs make this spectral
transition to/from an EUV bright accretion disc to a hot inner flow,
and that the change in BLR properties is a mostly a consequence of
the changing shape of the ionizing spectrum.

While the state transition describes the changing
shape of the spectrum, the timescale for disc evaporation/condensation should be the
viscous timescale of the disc. This predicts far too long a timescale compared to that observed.
There is a similar problem in explaining the background decline in flux which triggers the transition.
In Mrk~1018 the optical flux drops by more than a factor of $10$, requiring a
true change in mass accretion rate over a timescale of $\sim 10$~years, orders of magnitude shorter
than the viscous timescale. We discuss all the factors which change between AGN and binary discs
so as to properly scale between them, as sizes and scale heights should be different. However, one
key difference which has not been considered so far is that AGN discs are dominated by radiation pressure
even in standard disc models. Thus the sound speed, and hence the viscous speed, is much faster for 
a given gas temperature. Including magnetic pressure to stabilize the disc again increases the sound speed, 
making it feasible that there can be correspondence between the observations and theory. 
We suggest the Hydrogen ionization disc instability, where the heating/cooling front travels on 
a speed which is faster than the viscous timescale by a factor $H/R$, as the underlying mechanism 
for the flux drop. We speculate that this is the origin of the most extreme AGN variability. 
Proper modeling of this is required in order to predict the behaviour in more detail.

\section*{Acknowledgements}

We thank the referee for his/her careful read of this manuscript 
and valuable comments.
HN is supported by Program for Establishing a Consortium for the Development 
of Human Resources in Science and Technology, Japan Science and Technology Agency (JST). 
CD acknowledges the Science and Technology Facilities Council (STFC) 
through grant ST/P000541/1 for support. 
This work made use of data supplied by the UK Swift Science Data Centre at the
University of Leicester.







%
%
%
%

\bsp	
\label{lastpage}
\end{document}